\documentclass[prd,preprint,nofootinbib]{revtex4}

\usepackage{graphicx}
\usepackage{amssymb}
\usepackage{amsmath}

\begin{document}

\title{\Large \bf
 Solving Cosmological Problems of Supersymmetric Axion Models
in Inflationary Universe
 }

\author{Masahiro Kawasaki$^{a b}$ and Kazunori Nakayama$^{a}$}

\affiliation{%
$^a$ Institute for Cosmic Ray Research,
     University of Tokyo, Chiba 277-8582, Japan\\
$^b$ Institute for the Physics and Mathematics of the Universe, 
     University of Tokyo, Chiba 277-8582, Japan
}

\date{\today}

\begin{abstract}
We revisit inflationary cosmology of axion models in the light of recent
developments on the inflaton decay in supergravity.  We find that all
the cosmological difficulties, including gravitino, axino overproduction
and axionic isocurvature fluctuation, can be avoided if the saxion field
has large initial amplitude during inflation and decays before big-bang
nucleosynthesis.
\end{abstract}

\preprint{IPMU-08-0025}
\maketitle

\section{Introduction}

Although the standard model of the particle physics has achieved great
successes, it still has some theoretical problems.  One is the gauge
hierarchy problem, and another is the strong CP problem.  On the
otherhand, cosmological problems such as  dark matter and baryon
asymmetry of the universe cannot be explained within the framework of
the standard model.  These problems indicate that there is an underlying
physics beyond the standard model.

Supersymmetry (SUSY) is one of the best motivated candidates as physics
beyond the standard model to solve the gauge hierarchy problem.  On the
other hand, Peccei-Quinn (PQ) mechanism is a simple promising solution
to the strong CP problem \cite{Peccei:1977hh}.  As a consequence of the
breaking of PQ symmetry, the existence of a pseudo-Nambu-Goldstone
boson, axion, is predicted \cite{Kim:1986ax}.  Thus we are eager to
study the axion models within the framework of SUSY.  However, cosmology
of SUSY axion models is highly non-trivial.  It is well known that
gravitino, which is the fermionic superpartner of the graviton and has a
long lifetime, is produced during reheating after inflation and its
decay gives serious effect to thermal histry of the universe (this is
called gravitino problem).  In addition to the gravitino, there appear
other long-lived particles called saxion, the scalar partner of the
axion, and axino, the fermionic superpartner of the axion
\cite{Rajagopal:1990yx}.  Both of them have a potential to cause
cosmological disaster, which may be even more problematic than the usual
gravitino problem
\cite{Chun:1993vz,Covi:2001nw,Choi:2008zq,
Chun:2008rp,Asaka:1998xa,Kawasaki:2007mk}.

Furthermore, recently it is pointed out that gravitinos are also
produced non-thermally from the inflaton decay
\cite{Endo:2006zj,Dine:2006ii,Endo:2006tf,Kawasaki:2006gs,
Endo:2007ih,Endo:2007sz}.  Inclusion of such contributions makes the
gravitino problem much worse so that some inflation models may be
excluded depending on the mass of the gravitino.

This is not the end of the story.  The axion field induces an
isocurvature fluctuation \cite{Seckel:1985tj}, whose amplitude is
proportional to the Hubble scale during inflation.  Recent cosmological
observations are consistent with pure adiabatic one as the
initial density fluctuation of the universe \cite{Bean:2006qz},
and hence the fraction of the isocurvature component to the density
perturbation is constrained.  This excludes the high-scale inflation
models such as chaotic inflation or many hybrid inflation models.

Thus SUSY axion models seem to suffer from various cosmological
problems, which exclude many inflation models.  However, in this paper
we point out that a resolution to all the above difficulties is already
built in SUSY axion models themselves.  The saxion is naturally expected
to have a large initial amplitude of order of the reduced Planck scale
$M_P$ during inflation, and starts  coherent oscillation after
inflation.  Soon after reheating due to the inflaton decay,
the saxion dominates the universe since the energy density of the saxion
is almost comparable to the total energy density during the
inflaton-dominated era.  Finally, the saxion decays at later epoch
and reheats the universe again releasing huge entropy.  A remarkable
feature is that huge entropy produced by the saxion decay dilutes the
possibly harmful gravitinos and axinos.  This provides a solution to the
cosmological gravitino and axino problems.  Moreover, the amplitude of
the axionic isocurvature fluctuation is also significantly reduced for a
large initial amplitude of the saxion, which makes high-scale inflation
models such as chaotic inflation compatible with cosmological
constraints.

This paper is organized as follows.  In Sec.\ref{sec:problem} we
overview cosmological problems with the gravitino, axino and axion.  In
Sec.\ref{sec:inflation}, we show that entropy-production by the saxion
decay overcomes these cosmological difficulties.  We conclude in
Sec.\ref{sec:conclusion}.


\section{Cosmological Problems} \label{sec:problem}

In this section we briefly summarize cosmological constraints on the
abundances of long-lived particles, the gravitino, axino and axion,
which appear in SUSY axion models.

\subsection{Gravitino}

Gravitinos are produced in the early universe through scatterings of
particles in thermal bath.  For an unstable gravitino, its decay may
significantly affect Big-Bang nucleosynthesis
(BBN)~\cite{Kawasaki:1994af,Kawasaki:2004yh}.  If the gravitino is
stable, it may have too large contribution to the present matter density
of the universe~\cite{Moroi:1993mb}.  Thus the gravitino abundance is
constrained for nearly all  mass range.  The abundance is calculated
as~\cite{Bolz:2000fu,Kawasaki:2004yh}
\begin{equation}
\begin{split}
  Y_{3/2}^{\rm (TP)}\simeq &1.9\times 10^{-12}
  \left (1+ \frac{m_{\tilde g}^2}{3m_{3/2}^2} \right ) 
	 \left ( \frac{T_R}{10^{10}~\rm{GeV}} \right ) \\
  & \times \left [1+ 0.045\ln \left (\frac{T_R}{10^{10}~\rm{GeV}} 
         \right ) \right ] 
  \left [1- 0.028\ln \left (\frac{T_R}{10^{10}~\rm{GeV}}\right ) \right ],
  \label{gravitinoTP}
\end{split}
\end{equation}
where $m_{\tilde g}$ is the gluino mass and $T_R$ is the reheating
temperature of the universe defined as $T_R = (10/\pi^2
g_*)^{1/4}\sqrt{\Gamma_{\rm total}M_P}$.

Recently, it was pointed out that gravitinos are also produced directly
from the decay of the inflaton through supergravity effects
\cite{Endo:2006zj,Dine:2006ii,Endo:2006tf,Kawasaki:2006gs}.  Taking
account of the mixing of the inflaton with SUSY breaking field, the
decay rate of the inflaton into gravitino pair is written as
\begin{equation}
   \Gamma(\phi \to \psi_{3/2}\psi_{3/2}) 
   \simeq \frac{|\mathcal G_\phi|^2}{288\pi}
   \frac{m_\phi^5}{m_{3/2}^2M_P^2},
\end{equation}
where $m_\phi$ denotes the inflaton mass and $\mathcal G_\phi$ is the
effective coupling to the gravitino given in Ref.~\cite{Endo:2006tf}.
In this paper we assume a dynamical SUSY breaking scenario with
dynamical scale $\Lambda$.  If $m_\phi \ll m_Z$ where $m_Z$ denotes the
mass of the SUSY breaking field (hereafter for simplicity we assume $m_Z
\sim \Lambda$), the direct production process cannot be suppressed.  In
this case, the effective coupling is given by $|\mathcal G_\phi| \sim
3(\langle \phi \rangle/M_P)(m_{3/2}/m_\phi)$ and the decay rate is
estimated as~\cite{Endo:2006tf,Endo:2007sz}
\begin{equation}
  \Gamma(\phi \to \psi_{3/2}\psi_{3/2}) \simeq \frac{1}{32\pi}
  \left ( \frac{\langle \phi \rangle}{M_P} \right )^2
  \frac{m_\phi^3}{M_P^2}.
\end{equation}
This gives the gravitino abundance as
\begin{equation}
\begin{split}
   Y_{3/2}^{\rm (NTP)} &\simeq 
   2\frac{\Gamma(\phi \to \psi_{3/2} \psi_{3/2})}{\Gamma_{\rm total}}
   \frac{3T_R}{4m_\phi} \\
   & \simeq 7\times 10^{-11} 
   \left ( \frac{\langle \phi \rangle}{10^{15}~{\rm GeV}} \right )^2
   \left ( \frac{m_\phi}{10^{12}~{\rm GeV}} \right )^2
   \left ( \frac{T_R}{10^{6}~{\rm GeV}} \right )^{-1}.
   \label{gravitinoNTPd}
\end{split}
\end{equation}
If $m_\phi > \Lambda $, the direct decay of the inflaton into gravitino
pair can be suppressed if there are no couplings such as $\delta K \sim
|\phi|^2 zz$ in the K{$\ddot {\rm a}$}hler potential.  Instead, the
inflaton decays into hidden gauge sector through the anomaly effects,
and each hidden hadron eventually produces at least one gravitino
\cite{Endo:2007ih}.  The partial decay rate of the inflaton into hidden
gauge sector is given by
\begin{equation}
   \Gamma_{\rm anomaly} \simeq \frac{N_h \alpha_h^2}{256\pi^3}
   (\mathcal T_G^h-\mathcal T_R^h)^2 
   \left ( \frac{\langle \phi \rangle}{M_P} \right )^2
   \frac{m_\phi^3}{M_P^2},
\end{equation}
where $N_h$ is the number of generators and $\alpha_h$ is the gauge
coupling constant of the hidden gauge group.  $\mathcal T_G^h$ and
$\mathcal T_R^h$ are the Dynkin index of the adjoint representation and
matter fields in the representation of the dimension $d_R$,
respectively.  Then the abundance of non-thermally produced gravitinos
is given by
\begin{equation}
\begin{split}
   Y_{3/2}^{\rm (NTP)} &\simeq 
   2N_{3/2}\frac{\Gamma_{\rm anomaly}}{\Gamma_{\rm total}}
   \frac{3T_R}{4m_\phi} \\
   & \simeq 9\times 10^{-13}\epsilon 
   \left ( \frac{\langle \phi \rangle}{10^{15}~{\rm GeV}} \right )^2
   \left ( \frac{m_\phi}{10^{12}~{\rm GeV}} \right )^2
   \left ( \frac{T_R}{10^{6}~{\rm GeV}} \right )^{-1},
   \label{gravitinoNTP}
\end{split}
\end{equation}
where $\epsilon$ is $O(1)$ constant given by $\epsilon = N_{3/2}N_h
\alpha_h^2 (\mathcal T_G^h-\mathcal T_R^h)^2$ (here $N_{3/2}$ denotes
the averaged number of produced gravitinos per hidden hadron jet).

If the gravitino is unstable, photons or hadrons produced through the
decay of the gravitino may affect light element abundances synthesized
through BBN.  On the other hand, if the gravitino is stable, it
contributes to the present dark matter density.  Both set the upper
bound on the gravitino abundance and reheating temperature of the
universe.  Moreover, we can see that $Y_{3/2}^{\rm (TP)}$ is
proportional to $T_R$, while $Y_{3/2}^{\rm (NTP)}$ is proportional to
$T_R^{-1}$.  Thus lowering $T_R$ does not ameliorate the situation.
Also it was found that the reheating temperature of the universe is
bounded from below, due to the spontaneous decay processes of the
inflaton through top Yukawa coupling~\cite{Endo:2006qk}, as
\begin{equation}
    T_R \gtrsim 3~{\rm TeV} |y_t|
    \left ( \frac{228.75}{g_*(T_R)} \right )^{1/4}
    \left ( \frac{m_{\varphi}}{10^{12}~{\rm GeV}} \right )^{3/2}
    \left ( \frac{\langle \phi \rangle}{10^{15}~{\rm GeV}} \right ), 
    \label{lowTR}
\end{equation}
where $y_t$ is the top Yukawa coupling.  Including non-thermally
produced contribution, many inflation models are severely constrained.
For example, if the gravitino mass is $\sim 1$~TeV and its hadronic
branching ratio is $O(1)$, typical inflation models such as new, hybrid
and chaotic inflation models are excluded~\cite{Endo:2007sz}.

\subsection{Axino}

In SUSY axion model, the axion forms a supemultiplet which includes
scalar partner of the axion called saxion, and fermionic superpartner of
the axion called axino.  Both of them have long lifetime and may
significantly affect
cosmology~\cite{Chun:1993vz,Covi:2001nw,Choi:2008zq,
Chun:2008rp,Asaka:1998xa,Kawasaki:2007mk}.  Similar to the gravitino,
axinos are also produced through scatterings of the particles in thermal
bath.  The resulting abundance is calculated as~\cite{Covi:2001nw}
\begin{equation}
  Y_{\tilde a}
  \simeq 2.0 \times 10^{-7}g_s^6 
  \left ( \frac{F_a}{10^{12}{\rm\,GeV}} \right )^{-2} 
  \left ( \frac{T_R}{10^{6}{\rm\,GeV}} \right ),
  \label{axinoTP}
\end{equation}
where $g_s$ is the QCD gauge coupling constant.  This expression is
valid for $T_R \gtrsim 10$~TeV, and the abundance is suppressed for $T_R
\lesssim 1$~TeV because SUSY particles are not produced efficiently.
Although the axino mass is model dependent, it may have the mass of
order of the gravitino mass \cite{Chun:1995hc}.  Here we assume that the
axino mass is of the order of the gravitino mass.

From (\ref{axinoTP}), one can see that very low reheating temperature is
needed in order to avoid the overproduction of the axino if it is the
LSP.  This sets an upper bound on $T_R$ as $T_R \lesssim 1$~TeV for
$m_{\tilde a} \gtrsim 1$~GeV, and $T_R \lesssim 1$TeV$(1{\rm
GeV}/m_{\tilde a})$ for $m_{\tilde a} \lesssim 1$~GeV when $F_a =
10^{11}$~GeV.

\subsection{Axion}  \label{sec:axion}

The axion is a pseudo-Nambu-Goldstone boson associated with spontaneous
breaking of the PQ symmetry.  The axion is practically massless for $T
\gtrsim 1$~GeV due to the finite-temperature effect, and begins to
oscillate after it becomes massive for $T \lesssim 1$~GeV.  The
abundance of axion in the form of coherent oscillation is estimated as
\cite{Kolb:1990}
\begin{equation}
   \Omega_a h^2 \sim 0.2 
   \left ( \frac{F_a \theta_i^{1.7}}{10^{12}~{\rm GeV}} \right )^{1.18},
\end{equation}
where $\theta_i$ denotes the initial misalignment angle of the axion,
which is naturally expected to be $O(1)$ without fine-tuning.  Thus the
upper bound on the PQ scale is $F_a \lesssim 10^{12}$~GeV, although
late-time entropy production can relax this upper bound
\cite{Steinhardt:1983ia}.  On the other hand, astrophysical arguments
require that $F_a$ should be larger than $\sim 10^{10}$~GeV
\cite{Raffelt:1990yz}.

If the PQ symmetry is broken during or before the inflation, the quantum
fluctuation of the axion induces isocurvature fluctuation with magnitude
$\sim H_I/(\pi s_i)$ \cite{Seckel:1985tj}, where $s_i$ denotes the field
value of the saxion during inflation.  Recent cosmological observations
indicate that the ratio of the magnitude of isocurvature perturbation to
adiabatic one should be less than about
$0.3$~\cite{Bean:2006qz,Beltran:2006sq}.  This leads to the constraint
on the Hubble scale during inflation,
\begin{equation}
  H_I \lesssim 2\times 10^7~{\rm GeV} ~\theta_i ^{-1}
  \left ( \frac{\Omega_m h^2}{0.13} \right )
  \left ( \frac{s_i}{F_a} \right )
  \left ( \frac{F_a}{10^{12}~{\rm GeV}} \right )^{-0.18}. 
  \label{isocurv}
\end{equation}
Thus, high-scale inflation models are excluded unless $s_i$ is as large
as the Planck scale.  As a result, low-scale inflation models such as a
new inflation model were considered to be suitable for the axion
cosmology.  Note that if $H_I \gtrsim F_a$, the PQ symmetry may be
restored during inflation and hence isocurvature constraint does not
apply, although there may be problematic domain wall formation for
generic axion models.

\begin{figure}[t]
 \begin{center}
  \includegraphics[width=0.5\linewidth]{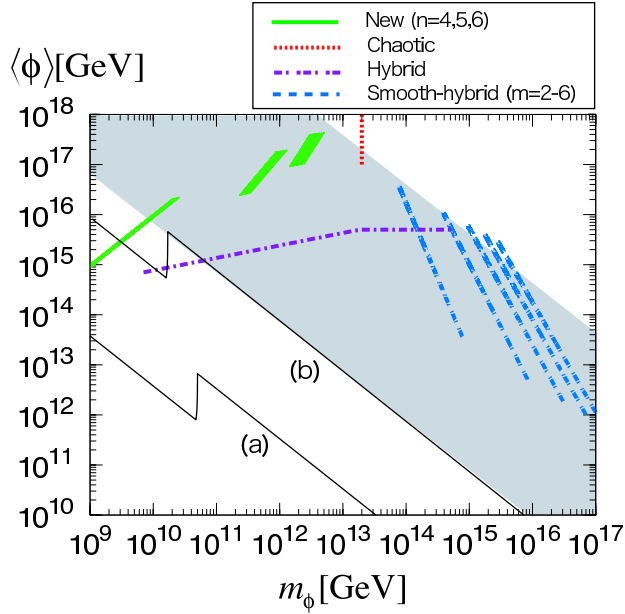} 
  \caption{ Axion isocurvature constraint on the $m_\phi$-$\langle \phi
  \rangle$ plane with $s_i=F_a=10^{12}~$GeV.  The shaded region is
  excluded from the isocurvature perturbation constraint.  Green solid
  lines represent new inflation models with $n=4,5,6$ from left to
  right, red dotted line represents chaotic inflation model, purple
  dot-dashed line represents hybrid-inflation model and blue dashed
  lines represent smooth-hybrid inflation models for $m=2$-$6$ from left
  to right (for definitions of $n$ and $m$ see footnote).  Region
  above thin solid black lines are excuded from gravitino overproduction
  in inflaton decay for (a) $m_{3/2}=1$~TeV and (b) $m_{3/2}=100$~GeV,
  with $T_R=1$~TeV.  }
  \label{fig:Si_Fa}
 \end{center}
\end{figure}

In Fig.~\ref{fig:Si_Fa} the axion isocurvature constraint is shown on  
the $m_\phi$-$\langle \phi \rangle$ plane with
$s_i=F_a=10^{12}~$GeV.\footnote{
Here we have estimated the inflation energy scale by $H_I \sim m_\phi
\langle \phi \rangle/\sqrt{3}M_P$.  For inflation models with the
potential like $V \sim (v^2 - g\phi^n)^2$, as is the case for new,
hybrid and smooth-hybrid inflation models, this estimation is correct
except for the numerical factor $n$.  For the chaotic inflation model,
this evaluation cannot be applied and the correct estimate is $H_I \sim
10^{14}~$GeV.}
Also we show the prediction of typical inflation models in supergravity,
new \cite{Kumekawa:1994gx,Asaka:1999jb},
hybrid \cite{Dvali:1994ms,BasteroGil:2006cm}, 
smooth-hybrid \cite{Lazarides:1995vr}, 
and chaotic inflation \cite{Kawasaki:2000yn} models.\footnote{
We assume the following superpotentials for new and smooth hybrid inflation models :
$W=\psi(v^2-g\phi^n)$ for new inflation and
$W=\psi[v^2-(\bar \phi \phi)^m/M^{2m-2}]$ for smooth hybrid new inflation model.
}
Furthermore, non-thermally produced gravitinos restrict the parameter
region.  Regions above thin solid black lines are excluded from
gravitino overproduction, for (a) $m_{3/2}=1$~TeV and (b)
$m_{3/2}=100$~GeV.  Here we set $T_R =1$~TeV, since axino thermal
production sets the upper bound on the reheating temperature as $T_R
\lesssim 1$~TeV.  Saxions are not harmful for cosmology for these
parameter sets.  We can see that most inflation models are excluded.
However, if one allows a large field value of the saxion during
inflation, the effective PQ scale during inflation can be practically
much larger than $F_a$, which suppress the magnitude of the isocurvature
perturbation~\cite{Linde:1991km}, as we will see in the next section.


\section{Saving Inflation Models}  \label{sec:inflation}

All the above cosmological difficulties are avoided by taking account of
the dynamics of the saxion.  In the above arguments, we have assumed
that there was no entropy production after reheating by the inflaton.
However, in SUSY axion model, the saxion can have the large initial
amplitude of order of the Planck scale. 
For example, let us consider the case where the saxion has negative 
Hubble mass term.
Since the saxion $s$ corresponds to a flat direction in the scalar 
potential, the saxion field rolls away due to the negative Hubble 
mass term during inflation until the field value becomes the Planck 
scale where the potential becomes steep as $\sim\exp(s^2/M_P^2)$.
Such saxion condensate provides a source of
late-time entropy production~\cite{Kim:1992eu} and dilutes the harmful
gravitino and axino.  Furthermore, large initial amplitude during
inflation suppresses the axionic isocurvature fluctuation.  Therefore,
there arises a possibility that many inflation models are free from
cosmological difficulties which we encountered in the previous section.

\subsection{Saxion with large initial amplitude}

The saxion corresponds to a flat direction along which the PQ scalars do
not feel the potential.  A flat direction is lifted by the SUSY breaking
effect and hence the saxion mass ($m_s$) is naturally expected to be of
order of the gravitino mass.

The coherent saxion oscillation starts at $H\sim m_s$, with the initial
amplitude $s_i$.  If $s_i \sim F_a$ and the saxion is relatively light,
its decay causes various cosmological difficulties, which severely
restricts the saxion abundance~\cite{Asaka:1998xa,Kawasaki:2007mk}.  As
a result, the upper bond on the reheating temperature becomes much more
stringent than the case where the saxion and axino are absent for wide
range of the saxion mass~\cite{Kawasaki:2007mk}.

Here we consider another possibility.  Thermal history of the universe
can be significantly modified due to the saxion coherent oscillation
with a large initial amplitude $s_i \sim M_P$.  In this case the saxion
dominates the universe immediately after the inflaton decays, and
finally the universe is reheated again by the saxion decay.  However, if
the saxion main decay mode is into two axions $(s\to 2a)$, the produced
axions contribute to the extra relativistic degrees of freedom, which
changes the Hubble expansion rate and the BBN prediction.  Thus we need
to investigate the saxion decay modes in order to ensure that the saxion
decay does not produce too many axions.  The decay rate into two axions
is estimated as~\cite{Chun:1995hc}
\begin{equation}
  \Gamma(s\to 2a) \simeq \frac{f^2}{64\pi}\frac{m_s^3}{F_a^2},
\end{equation}
where $f=\sum_i q_i^3 v_i^2/F_a^2$ with the VEV of the $i$-th PQ scalar
field $v_i$ and its PQ charge $q_i$.  This often gives the dominant
contribution to the total saxion decay rate if $f \sim 1$.  In order to
realize consistent cosmology, this must not be the main decay mode.

Another important decay mode is into gluons, arising from the coupling
through the QCD anomaly effect.  The decay rate is estimated as
\begin{equation}
  \Gamma_s(s\to gg) \simeq \frac{\alpha_s^2}{32\pi^3}\frac{m_s^3}{F_a^2}.
\end{equation}
For the DFSZ axion model~\cite{Zhitnitsky:1980tq}, the saxion can also
decay into fermion-anti-fermion pair through the coupling arising from
mixing of the PQ scalar with MSSM Higgs doublets.  The decay rate is
estimated as
\begin{equation}
  \Gamma(s\to d_i \bar d_i) \simeq \frac{3}{8\pi} 
  \left ( \frac{2x}{x+x^{-1}} \right )^2
  m_s \left ( \frac{m_{di}}{F_a} \right )^2
  \left (1- \frac{4m_{di}^2}{m_s^2} \right )^{3/2},
\end{equation}
for the decay into down-type quarks $d_i$ ($i=1,2,3$) where $x=\tan
\beta =\langle H_u \rangle/ \langle H_d \rangle$, and
\begin{equation}
  \Gamma(s\to u_i \bar u_i) \simeq  \frac{3}{8\pi} 
  \left ( \frac{2x^{-1}}{x+x^{-1}} \right )^2
  m_s \left ( \frac{m_{ui}}{F_a} \right )^2 
  \left (1- \frac{4m_{ui}^2}{m_s^2} \right )^{3/2}.
\end{equation}
for the decay into up-type quarks $u_i$ ($i=1,2,3$).
The decay rate into Higgs boson pair is also comparable, 
\begin{equation}
	\Gamma(s\to hh) \simeq  \frac{1}{8\pi}\frac{m_s^3}{F_a^2}
	 \left ( \frac{\mu}{m_s} \right )^4
	 \left (1- \frac{4m_{h}^2}{m_s^2} \right )^{1/2},
\end{equation}
where $\mu$-parameter is of order of the weak scale and set to $300$~GeV
here.  For simplicity we consider only the decay into the lightest Higgs
bosons.  Decays into other Higgs bosons also have comparable rate if
kinematically allowed.  Also we neglect decay into SUSY particle pair
assuming that such decay modes are kinematically forbidden.

In order not to contradict with observations, the increase of the
effective number of neutrino species $\Delta N_\nu$ should be smaller
than about 1, which restricts the branching ratio into axions $B_a$ as
\begin{equation}
  \frac{B_a}{1-B_a} \lesssim \frac{7}{43}
  \left ( \frac{g_*(T_s)}{10.75} \right )^{1/3} 
  \Delta N_{\nu ({\rm bound})},
  \label{Babound}
\end{equation}
where $\Delta N_{\nu ({\rm bound})} \sim 1$, and $T_s$ denotes the decay
temperature of the saxion.

\begin{figure}[tbp]
 \begin{center}
  \includegraphics[width=0.5\linewidth]{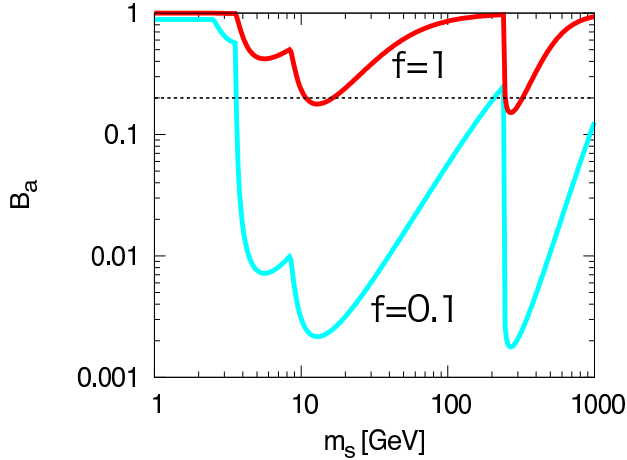} 
  \caption{ Branching ratio into axions in DFSZ model for $f=1$ (upper)
  and $f=0.1$ (lower).  Region above the thin dashed line is excluded.  }
  \label{fig:Ba}
 \end{center}
\end{figure}

In Fig.~\ref{fig:Ba}, $B_a$ as a function of the saxion mass with $f=1$
and $f=0.1$ is shown.  As a representative value, we show $B_a = 0.2$
by the thin dashed line above which too many axions are produced by the
saxion decay and the constraint (\ref{Babound}) is not satisfied.  It
can be seen that for some mass ranges, the decay into axions is
subdominant process.  For that case the saxion is an ideal
candidate of the source of late-time entropy production.


\subsection{Dilution of gravitinos and axinos}

We have shown that late-time entropy production by the saxion decay can
can take place without producing too many axions.  Now let us
investigate the gravitino and axino abundances after the dilution by the
saxion decay.

\begin{figure}[htbp]
 \begin{center}
   \includegraphics[width=0.5\linewidth]{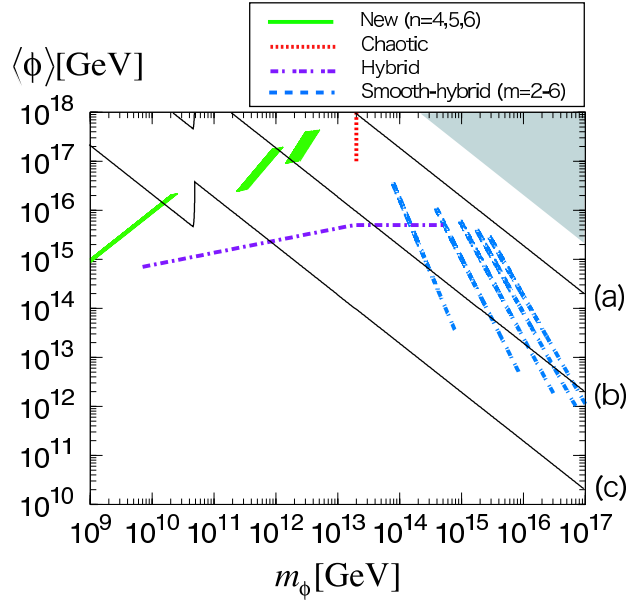} 
  \caption{ Constraints on inflation models for $m_{3/2}=1$~TeV and
  $F_a=10^{12}$~GeV with $s_i=M_P$.  Region above the thin solid black
  lines is excluded for (a) $T_R=10^9$~GeV, (b) $T_R=10^7$~GeV and (c)
  $T_R=10^5$~GeV.  The shaded region is excluded from axionic
  isocurvature constraint.  } 
  \label{fig:1TeV}
 \end{center}
\end{figure}

\begin{figure}[htbp]
 \begin{center}
   \includegraphics[width=0.5\linewidth]{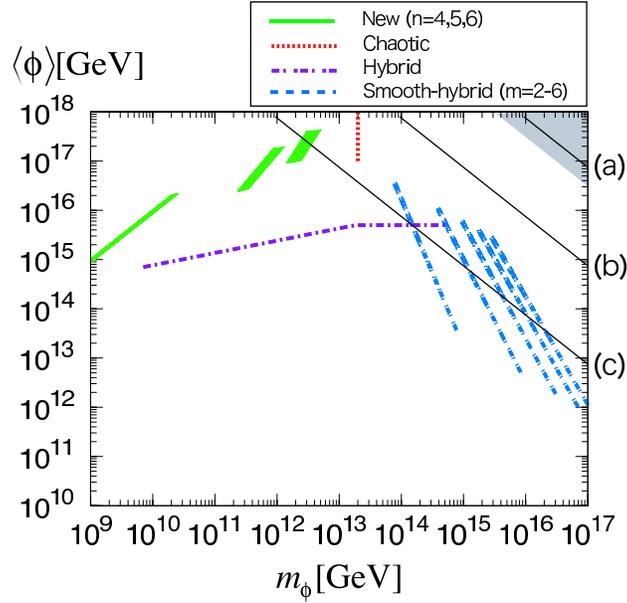} 
  \caption{ Same as Fig.~\ref{fig:1TeV}, except for $m_{3/2}=100$~GeV
  and $F_a=10^{11}$~GeV.}  
  \label{fig:100GeV}
 \end{center}
\end{figure}

If we assume the main decay mode is $s\to b \bar b$,
the typical decay temperature of the saxion is estimated as
\begin{equation}
   T_s \sim 250~{\rm MeV}\left ( \frac{g_*(T_s)}{10} \right )^{-1/4}
   \left ( \frac{m_s}{100~{\rm GeV}} \right )^{1/2}
   \left ( \frac{F_a}{10^{11}~{\rm GeV}} \right )^{-1},
\end{equation}
which is much lower than the typical reheating temperature after
inflation.  Thus, possibly harmful axinos and gravitinos, produced
either thermally or non-thermally, are diluted by the saxion decay.
After the dilution, thermally produced gravitinos during reheating
processes after inflation have only negligible abundance of order
$Y_{3/2}^{({\rm TP})} \lesssim 10^{-20}$ independently of $T_R$.  On the
other hand, the gravitino abundance produced from inflaton decay is
estimated as
\begin{equation}
\begin{split}
   Y_{3/2}^{\rm (NTP)} &= 
   \frac{9\Gamma(\phi\to\psi_{3/2}\psi_{3/2})}{2\Gamma_{\rm total}}
   \frac{T_s}{m_\phi}\left ( \frac{M_P}{s_i} \right )^2 \gamma \\
   &\sim 2.0\times 10^{-12} \gamma
   \left ( \frac{T_s}{1~{\rm GeV}} \right )
   \left ( \frac{T_R}{10^7~{\rm GeV}} \right )^{-2}
   \left ( \frac{m_\phi}{10^{15}~{\rm GeV}} \right )^{2}
   \left ( \frac{\langle \phi \rangle}{10^{15}~{\rm GeV}} \right )^{2}
   \left ( \frac{M_P}{s_i} \right )^{2},
\end{split}
\end{equation}
for $m_\phi < \Lambda$, and
\begin{equation}
\begin{split}
   Y_{3/2}^{\rm (NTP)} &= 
   \frac{9N_{3/2}\Gamma_{\rm anomaly}}{2\Gamma_{\rm total}}
   \frac{T_s}{m_\phi}\left ( \frac{M_P}{s_i} \right )^2 \gamma \\
   &\sim 2.6\times 10^{-14}\epsilon \gamma
   \left ( \frac{T_s}{1~{\rm GeV}} \right )
   \left ( \frac{T_R}{10^7~{\rm GeV}} \right )^{-2}
   \left ( \frac{m_\phi}{10^{15}~{\rm GeV}} \right )^{2}
   \left ( \frac{\langle \phi \rangle}{10^{15}~{\rm GeV}} \right )^{2}
   \left ( \frac{M_P}{s_i} \right )^{2},
\end{split}
\end{equation}
for $m_\phi > \Lambda$ where $\gamma$ is defined as
\begin{equation}
   \gamma = \left \{
	     \begin{array}{ll}
	      1  &{\rm for}~~~T_{\rm osc} > T_R \\
	      T_R/T_{\rm osc} &{\rm for}~~~T_{\rm osc} < T_R
	     \end{array}
	      \right. ,
\end{equation}
with the temperature at which the saxion starts to oscillate $T_{\rm
osc}$.  Axinos are also diluted by the saxion decay and the resultant
abundance is estimated as\footnote{
Here we have assumed that the reheating due to the inflaton decay
completes after the saxion oscillation. Otherwise, the axino
abundance is suppressed by $s_i^2 \sim M_P^2$, not $F_a^2$, and is
safely neglected.  }
\begin{equation}
 Y_{\tilde a}
  \simeq 6.0 \times 10^{-13}g_s^6 
  \left ( \frac{F_a}{10^{12}{\rm\,GeV}} \right )^{-2} 
  \left ( \frac{T_s}{1{\rm\,GeV}} \right )
  \left ( \frac{M_P}{s_i} \right )^{2},
  \label{axinoTPdil}
\end{equation}
In Figs.~\ref{fig:1TeV} and \ref{fig:100GeV}, the allowed and excluded
region in $m_\phi$-$\langle \phi \rangle$ plane are shown for the
unstable gravitino with $m_{3/2}=1$~TeV and stable gravitino with
$m_{3/2}=100$~GeV.  We put a constraint $Y_{3/2}< 10^{-16}$ for the
former case \cite{Kohri:2005wn}.  Note that axinos decay well before BBN
for $m_{\tilde a}\gtrsim 1$~TeV and do not affect BBN.  Instead, the LSP
abundance emitted from the axino puts a constraint on the axino
abundance, and this is also satisfied.  It can be seen that all of the
inflation models are allowed for the reheating temperature $T_R \gtrsim
10^9 (10^6)$~GeV for $m_{3/2}= 1$~TeV (100~GeV).  Moreover, the
constraint on the inflation scale (\ref{isocurv}) is relaxed for $s_i
\sim M_P$, which enables even the chaotic inflation model to be
consistent with observations.\footnote{
Again it should be noted that isocurvature constraint on the chaotic
inflation model cannot be read from these figures. See the footnote in
Sec.\ref{sec:axion}.  }


Thus gravitinos are sufficiently diluted so that they do not cause
cosmological difficulties.  Interestingly, for the gravitino mass
$m_{3/2}\sim100$~GeV, non-thermally produced gravitinos can have the
desired abundance as dark matter of the universe.  One issue to be
addressed is the free-streaming length of the gravitino dark matter.  It
is calculated as
\begin{equation}
   \lambda_{\rm FS} = 
   \int_{t_i}^{t_{\rm eq}}\frac{v(t)}{a(t)}dt
   \sim R_{\rm eq}u_{\rm eq}\ln 
   \left [ \frac{1}{u_{\rm eq}} + \sqrt{ 1+\frac{1}{u_{\rm eq}^2} } \right ],
\end{equation}
where $R_{\rm eq}$ is the comoving Hubble scale ($\sim 108$Mpc)
and $u_{\rm eq} = v/\sqrt{1-v^2}$ with the velocity of the gravitino $v$,
both are evaluated at the time of matter-radiation equality.
$u_{\rm eq}$ is estimated as
\begin{equation}
   u_{\rm eq} \sim 8\times 10^{-9}
   \left ( \frac{m_\phi}{10^{15}~{\rm GeV}} \right )
   \left ( \frac{m_{3/2}}{1~{\rm TeV}} \right )^{-1}
   \left ( \frac{T_s}{1~{\rm GeV}} \right )^{1/3}
   \left ( \frac{T_R}{10^8~{\rm GeV}} \right )^{-4/3}.
\end{equation}
Hence the free-streaming length is negligibly small and the gravitino
dark matter acts as cold dark matter.  The axion is diluted by the
saxion decay because $T_s \lesssim 1$~GeV for $m_s \lesssim 100$~GeV,
but it may still have the comparable abundance to the present dark
matter abundance.

For $m_s \gtrsim 1$~TeV, there arises a possibility that the saxion
decays into SUSY particles if the decay is kinematically allowed.
According to Ref.~\cite{Endo:2006ix}, the decay rate of the saxion into
gauginos are roughly the same as that into gauge bosons.  Thus in
general LSPs are overproduced by the saxion decay.  However, for the LSP
with rather large annihilation cross section, these non-thermally
produced LSP abundance is reduced.  The abundance is given by
\begin{equation}
 \frac{\rho_{\rm LSP}}{s} \simeq 
  {\rm min} \left [ 
	     B_s\frac{2m_{\rm LSP}}{m_s} \frac{\rho_s}{s},~
	     \sqrt{\frac{45}{8\pi^2 g_*(T_s)} } 
	     \frac{m_{\rm LSP}}{\langle \sigma v \rangle T_s M_P}
            \right ] 
	    \label{NTLSP},
\end{equation}
where $B_s$ denotes the branching ratio of the saxion into SUSY
particles, and $\langle \sigma v \rangle$ denotes the thermally averaged
annihilation cross section of the LSP.  For the LSP with $\langle \sigma
v \rangle \sim 10^{-7}~{\rm GeV}^{-2}$, as is realized for the case of
wino- or higgsino-like LSP with the mass of
$O(100)$~GeV~\cite{Moroi:1999zb}, or bino-like LSP in the $S$-channel
resonance region~\cite{Nagai:2007ud}, the resultant LSP abundance can
account for the present dark matter of the universe.  Also there is 
contribution from the axion coherent oscillation to the dark matter
abundance.  Thus the dark matter may consist of a mixture of the axions
and non-thermally produced LSPs.

\subsection{Baryon asymmetry}

Note that baryon asymmetry is also diluted by the saxion decay.  Here we
show that Affleck-Dine mechanism \cite{Affleck:1984fy} can create 
appropriate amount of baryon asymmetry.\footnote{
See e.g., Ref.~\cite{Kawasaki:2007yy} for more details
in the case of late-time entropy production.}
Let us denote the Affleck-Dine (AD) field, which is one of the flat
directions in the scalar potential in the MSSM, as $\psi$.  The flat
direction is lifted by SUSY breaking effects and non-renormalizable
terms in the superpotential $W_{\rm NR}= \psi^n/nM^{n-3}$ where $n(\geq
4)$ is an integer and $M$ denotes the cutoff scale.  The resultant
baryon asymmetry created by the coherent motion of the AD field is
estimated as
\begin{equation}
\begin{split}
   \frac{n_B}{s}=\frac{n_B}{\rho_s}\frac{\rho_s}{s} 
   \simeq \frac{\delta_{\rm CP}m_{3/2}|\psi_{\rm os}|^2}{m_\psi^2 s_i^2}
   \frac{3T_s}{4},
\end{split}
\end{equation}
where $\delta_{\rm CP}$ denotes the effective CP angle, which is
naturally expected to be $O(1)$, and $\psi_{\rm os}$ is the field value
at the onset of the oscillation of the AD field.  If there exists a
negative Hubble mass term for the AD field, the field value is given by
$|\psi_{\rm os}|\sim (m_\psi M^{n-3})^{1/(n-2)}$.  For the specific case
$n=6$, we obtain
\begin{equation}
	\frac{n_B}{s} \sim 2\times 10^{-11} \delta_{\rm CP}
	\left ( \frac{m_\psi}{1~{\rm TeV}} \right )^{-3/2}
	\left ( \frac{m_{3/2}}{1~{\rm TeV}} \right )
	\left ( \frac{T_s}{1~{\rm GeV}} \right )
	\left ( \frac{M}{M_P} \right )^{3/2}
	\left ( \frac{M_P}{s_i} \right )^{2},
\end{equation}
and hence AD mechanism works well.  Although relatively large Q-balls
are formed through the AD mechanism ($Q\sim 10^{20}$)
\cite{Kusenko:1997zq}, 
their
cosmological effects are safely neglected, since Q-balls decay before
the saxion decays and LSPs emitted non-thermally by the Q-ball decay is
diluted by the entropy-production from the saxion.\footnote{
The Affleck-Dine mechanism described above has the baryonic isocurvature 
problem if the Hubble induced $A$-term is absent as expected for 
most inflation models~\cite{Kasuya:2008}. 
In this case high scale infaltion models are disfavored. However, the problem is avoided
for AD mechanism without superpotential~\cite{Kawasaki:2007yy}. }

\section{Conclusions }  \label{sec:conclusion}

We have shown that in SUSY axion models the saxion decay naturally
dilutes the axino and gravitino abundances produced both thermally or
non-thermally if the initial amplitude is of the order of $M_P$.  The
large initial amplitude of the saxion also saves high-scale inflation
models such as chaotic inflation and hybrid inflation models from
producing too much axionic isocurvature fluctuation.  Note that the
preexisting baryon asymmetry is also diluted by the saxion.  But the
Affleck-Dine mechanism can create large baryon asymmetry which survives
the dilution.

Finally we mention another possibility to dilute the gravitino and axino
abundances.  One may consider that thermal
inflation~\cite{Yamamoto:1985rd,Lyth:1995hj} can provide sufficient
dilution.  If the PQ scalar is trapped at the origin during inflation
and remains there until later epoch due to the finite-temperature
effect, the PQ scalar itself can cause a thermal inflation for some
class of models~\cite{Choi:1996vz}.  Axionic isocurvature fluctuation
does not arise since the PQ symmetry is restored during inflation.
However, we must take care of topological defects formation after the
onset of the PQ scalar oscillation, which may spoil the subsequent
cosmological evolution scenario.

\section*{Acknowledgment}

K.N. would like to thank the Japan Society for the Promotion of
Science for financial support.  This work was supported in part by the
Grant-in-Aid for Scientific Research from the Ministry of Education,
Science, Sports, and Culture of Japan, No. 18540254 and No 14102004
(M.K.).  This work was also supported in part by JSPS-AF Japan-Finland
Bilateral Core Program (M.K.).
This work was also supported by World Premier International
Research Center InitiativeiWPI Initiative), MEXT, Japan.

{}


\begin{thebibliography}{90}


\bibitem{Peccei:1977hh}
  R.~D.~Peccei and H.~R.~Quinn,
  Phys.\ Rev.\ Lett.\  {\bf 38}, 1440 (1977).
 
 
\bibitem{Kim:1986ax}
  For a review, see J.~E.~Kim,
  Phys.\ Rept.\  {\bf 150}, 1 (1987).
  

\bibitem{Rajagopal:1990yx}
  K.~Rajagopal, M.~S.~Turner and F.~Wilczek,
  Nucl.\ Phys.\ B {\bf 358}, 447 (1991).


\bibitem{Chun:1993vz}
  E.~J.~Chun, H.~B.~Kim and J.~E.~Kim,
  Phys.\ Rev.\ Lett.\  {\bf 72}, 1956 (1994)
  [arXiv:hep-ph/9305208];
  L.~Covi, J.~E.~Kim and L.~Roszkowski,
  Phys.\ Rev.\ Lett.\  {\bf 82}, 4180 (1999)
  [arXiv:hep-ph/9905212].
  

\bibitem{Covi:2001nw}
  L.~Covi, H.~B.~Kim, J.~E.~Kim and L.~Roszkowski,
  JHEP {\bf 0105}, 033 (2001)
  [arXiv:hep-ph/0101009];
  A.~Brandenburg and F.~D.~Steffen,
  JCAP {\bf 0408}, 008 (2004)
  [arXiv:hep-ph/0405158].
  
  
\bibitem{Choi:2008zq}
  K.~Y.~Choi, J.~E.~Kim, H.~M.~Lee and O.~Seto,
  arXiv:0801.0491 [hep-ph].
  
  
\bibitem{Chun:2008rp}
  E.~J.~Chun, H.~B.~Kim, K.~Kohri and D.~H.~Lyth,
  arXiv:0801.4108 [hep-ph].
  
  
\bibitem{Asaka:1998xa}
  T.~Asaka and M.~Yamaguchi,
  Phys.\ Rev.\  D {\bf 59}, 125003 (1999)
  [arXiv:hep-ph/9811451].
  
    
\bibitem{Kawasaki:2007mk}
  M.~Kawasaki, K.~Nakayama and M.~Senami,
  arXiv:0711.3083 [hep-ph].
  
  
\bibitem{Endo:2006zj}
  M.~Endo, K.~Hamaguchi and F.~Takahashi,
  Phys.\ Rev.\ Lett.\  {\bf 96}, 211301 (2006)
  [arXiv:hep-ph/0602061];
  S.~Nakamura and M.~Yamaguchi,
  Phys.\ Lett.\  B {\bf 638}, 389 (2006)
  [arXiv:hep-ph/0602081].


\bibitem{Dine:2006ii}
  M.~Dine, R.~Kitano, A.~Morisse and Y.~Shirman,
  Phys.\ Rev.\  D {\bf 73}, 123518 (2006)
  [arXiv:hep-ph/0604140].
  

\bibitem{Endo:2006tf}
  M.~Endo, K.~Hamaguchi and F.~Takahashi,
  Phys.\ Rev.\  D {\bf 74}, 023531 (2006)
  [arXiv:hep-ph/0605091].


\bibitem{Kawasaki:2006gs}
  M.~Kawasaki, F.~Takahashi and T.~T.~Yanagida,
  Phys.\ Lett.\  B {\bf 638}, 8 (2006)
  [arXiv:hep-ph/0603265];
  Phys.\ Rev.\  D {\bf 74}, 043519 (2006)
  [arXiv:hep-ph/0605297].


\bibitem{Endo:2007ih}
  M.~Endo, F.~Takahashi and T.~T.~Yanagida,
  Phys.\ Lett.\  B {\bf 658}, 236 (2008)
  [arXiv:hep-ph/0701042].

  
\bibitem{Endo:2007sz}
  M.~Endo, F.~Takahashi and T.~T.~Yanagida,
  Phys.\ Rev.\  D {\bf 76}, 083509 (2007)
  [arXiv:0706.0986 [hep-ph]].
  
  
\bibitem{Seckel:1985tj}
  D.~Seckel and M.~S.~Turner,
  Phys.\ Rev.\  D {\bf 32}, 3178 (1985);
  M.~S.~Turner and F.~Wilczek,
  Phys.\ Rev.\ Lett.\  {\bf 66}, 5 (1991).
  
    
\bibitem{Bean:2006qz}
  R.~Bean, J.~Dunkley and E.~Pierpaoli,
  Phys.\ Rev.\  D {\bf 74}, 063503 (2006)
  [arXiv:astro-ph/0606685];
  R.~Trotta,
  Mon.\ Not.\ Roy.\ Astron.\ Soc.\ Lett.\  {\bf 375}, L26 (2007)
  [arXiv:astro-ph/0608116];
  R.~Keskitalo, H.~Kurki-Suonio, V.~Muhonen and J.~Valiviita,
  JCAP {\bf 0709}, 008 (2007)
  [arXiv:astro-ph/0611917];
  M.~Kawasaki and T.~Sekiguchi,
  arXiv:0705.2853 [astro-ph].
  
  
\bibitem{Beltran:2006sq}
  M.~Beltran, J.~Garcia-Bellido and J.~Lesgourgues,
  Phys.\ Rev.\  D {\bf 75}, 103507 (2007)
  [arXiv:hep-ph/0606107].
  
      
  \bibitem{Kawasaki:1994af}
  M.~Kawasaki and T.~Moroi,
  Prog.\ Theor.\ Phys.\  {\bf 93}, 879 (1995)
  [arXiv:hep-ph/9403364];
  Astrophys.\ J.\  {\bf 452}, 506 (1995)
  [arXiv:astro-ph/9412055];
    E.~Holtmann, M.~Kawasaki, K.~Kohri and T.~Moroi,
    Phys.\ Rev.\ D {\bf 60}, 023506 (1999)
    [arXiv:hep-ph/9805405];
  K.~Jedamzik,
  Phys.\ Rev.\ Lett.\  {\bf 84}, 3248 (2000)
  [arXiv:astro-ph/9909445];
    M.~Kawasaki, K.~Kohri and T.~Moroi,
    Phys.\ Rev.\ D {\bf 63}, 103502 (2001)
    [arXiv:hep-ph/0012279];
    K.~Kohri,
    Phys.\ Rev.\ D {\bf 64}, 043515 (2001)
    [arXiv:astro-ph/0103411];
    R.~H.~Cyburt, J.~R.~Ellis, B.~D.~Fields and K.~A.~Olive,
    Phys.\ Rev.\ D {\bf 67}, 103521 (2003)
    [arXiv:astro-ph/0211258];
  K.~Jedamzik,
  Phys.\ Rev.\  D {\bf 74}, 103509 (2006)
  [arXiv:hep-ph/0604251].
    
 
 \bibitem{Kawasaki:2004yh}
   M.~Kawasaki, K.~Kohri and T.~Moroi,
   Phys.\ Lett.\ B {\bf 625}, 7 (2005)
   [arXiv:astro-ph/0402490];
   Phys.\ Rev.\ D {\bf 71}, 083502 (2005)
   [arXiv:astro-ph/0408426].
  
  
\bibitem{Moroi:1993mb}
  T.~Moroi, H.~Murayama and M.~Yamaguchi,
  Phys.\ Lett.\  B {\bf 303}, 289 (1993).
  
  
\bibitem{Bolz:2000fu}
  M.~Bolz, A.~Brandenburg and W.~Buchmuller,
  Nucl.\ Phys.\  B {\bf 606}, 518 (2001)
  [arXiv:hep-ph/0012052];
  J.~Pradler and F.~D.~Steffen,
  Phys.\ Rev.\  D {\bf 75}, 023509 (2007)
  [arXiv:hep-ph/0608344];
  Phys.\ Lett.\  B {\bf 648}, 224 (2007)
  [arXiv:hep-ph/0612291];
   V.~S.~Rychkov and A.~Strumia,
  Phys.\ Rev.\  D {\bf 75}, 075011 (2007)
  [arXiv:hep-ph/0701104].
  
  
\bibitem{Endo:2006qk}
  M.~Endo, M.~Kawasaki, F.~Takahashi and T.~T.~Yanagida,
  Phys.\ Lett.\  B {\bf 642}, 518 (2006)
  [arXiv:hep-ph/0607170].

    
\bibitem{Chun:1995hc}
  E.~J.~Chun and A.~Lukas,
  Phys.\ Lett.\  B {\bf 357}, 43 (1995)
  [arXiv:hep-ph/9503233].
    
  
\bibitem{Kolb:1990}
  E.~W.~Kolb and M.~S.~Turner, {\it The Early Universe},
  (Addison-Wesley, Reading, MA, 1990).
  
  
\bibitem{Steinhardt:1983ia}
  P.~J.~Steinhardt and M.~S.~Turner,
  Phys.\ Lett.\  B {\bf 129}, 51 (1983);
  G.~Lazarides, C.~Panagiotakopoulos and Q.~Shafi,
  Phys.\ Lett.\  B {\bf 192}, 323 (1987);
  G.~Lazarides, R.~K.~Schaefer, D.~Seckel and Q.~Shafi,
  Nucl.\ Phys.\  B {\bf 346}, 193 (1990);
  M.~Kawasaki, T.~Moroi and T.~Yanagida,
  Phys.\ Lett.\  B {\bf 383}, 313 (1996)
  [arXiv:hep-ph/9510461].
  
  
\bibitem{Raffelt:1990yz}
  G.~G.~Raffelt,
  Phys.\ Rept.\  {\bf 198}, 1 (1990).
    
\bibitem{Kumekawa:1994gx}
  K.~Kumekawa, T.~Moroi and T.~Yanagida,
  Prog.\ Theor.\ Phys.\  {\bf 92}, 437 (1994)
  [arXiv:hep-ph/9405337];
  K.~I.~Izawa and T.~Yanagida,
  Phys.\ Lett.\  B {\bf 393}, 331 (1997)
  [arXiv:hep-ph/9608359];
  M.~Ibe, K.~I.~Izawa, Y.~Shinbara and T.~T.~Yanagida,
  Phys.\ Lett.\  B {\bf 637}, 21 (2006)
  [arXiv:hep-ph/0602192].


  
  
\bibitem{Asaka:1999jb}
  T.~Asaka, K.~Hamaguchi, M.~Kawasaki and T.~Yanagida,
  Phys.\ Rev.\  D {\bf 61}, 083512 (2000)
  [arXiv:hep-ph/9907559];
  V.~N.~Senoguz and Q.~Shafi,
  Phys.\ Lett.\  B {\bf 596}, 8 (2004)
  [arXiv:hep-ph/0403294];


\bibitem{Dvali:1994ms}
  G.~R.~Dvali, Q.~Shafi and R.~K.~Schaefer,
  Phys.\ Rev.\ Lett.\  {\bf 73}, 1886 (1994)
  [arXiv:hep-ph/9406319];
  A.~D.~Linde and A.~Riotto,
  Phys.\ Rev.\  D {\bf 56}, 1841 (1997)
  [arXiv:hep-ph/9703209].


\bibitem{BasteroGil:2006cm}
  M.~Bastero-Gil, S.~F.~King and Q.~Shafi,
  Phys.\ Lett.\  B {\bf 651}, 345 (2007)
  [arXiv:hep-ph/0604198];
  M.~ur Rehman, V.~N.~Senoguz and Q.~Shafi,
  Phys.\ Rev.\  D {\bf 75}, 043522 (2007)
  [arXiv:hep-ph/0612023].
  


\bibitem{Lazarides:1995vr}
  G.~Lazarides and C.~Panagiotakopoulos,
  Phys.\ Rev.\  D {\bf 52}, 559 (1995)
  [arXiv:hep-ph/9506325].
  

  
  
\bibitem{Kawasaki:2000yn}
  M.~Kawasaki, M.~Yamaguchi and T.~Yanagida,
  Phys.\ Rev.\ Lett.\  {\bf 85}, 3572 (2000)
  [arXiv:hep-ph/0004243].


\bibitem{Linde:1991km}
  A.~D.~Linde,
  Phys.\ Lett.\  B {\bf 259}, 38 (1991).
  
  
\bibitem{Kim:1992eu}
  J.~E.~Kim,
  Phys.\ Rev.\ Lett.\  {\bf 67}, 3465 (1991);
  D.~H.~Lyth,
  Phys.\ Rev.\  D {\bf 48}, 4523 (1993)
  [arXiv:hep-ph/9306293];
  M.~Hashimoto, K.~I.~Izawa, M.~Yamaguchi and T.~Yanagida,
  Phys.\ Lett.\  B {\bf 437}, 44 (1998)
  [arXiv:hep-ph/9803263];
  T.~Banks, M.~Dine and M.~Graesser,
  Phys.\ Rev.\  D {\bf 68}, 075011 (2003)
  [arXiv:hep-ph/0210256].
  
    
\bibitem{Zhitnitsky:1980tq}
  A.~R.~Zhitnitsky,
  Sov.\ J.\ Nucl.\ Phys.\  {\bf 31} (1980) 260;
  M.~Dine, W.~Fischler and M.~Srednicki,
  Phys.\ Lett.\ B {\bf 104}, 199 (1981).
    
  
\bibitem{Kohri:2005wn}
  K.~Kohri, T.~Moroi and A.~Yotsuyanagi,
  Phys.\ Rev.\  D {\bf 73}, 123511 (2006)
  [arXiv:hep-ph/0507245].
  

\bibitem{Endo:2006ix}
  M.~Endo and F.~Takahashi,
  Phys.\ Rev.\  D {\bf 74}, 063502 (2006)
  [arXiv:hep-ph/0606075].


\bibitem{Moroi:1999zb}
  T.~Moroi and L.~Randall,
  Nucl.\ Phys.\  B {\bf 570}, 455 (2000)
  [arXiv:hep-ph/9906527];
  M.~Fujii and K.~Hamaguchi,
  Phys.\ Lett.\  B {\bf 525}, 143 (2002)
  [arXiv:hep-ph/0110072];
  Phys.\ Rev.\  D {\bf 66}, 083501 (2002)
  [arXiv:hep-ph/0205044].
  

\bibitem{Nagai:2007ud}
  M.~Nagai and K.~Nakayama,
  Phys.\ Rev.\  D {\bf 76}, 123501 (2007)
  [arXiv:0709.3918 [hep-ph]].
  
    
\bibitem{Affleck:1984fy}
  I.~Affleck and M.~Dine,
  Nucl.\ Phys.\  B {\bf 249}, 361 (1985);
  M.~Dine, L.~Randall and S.~D.~Thomas,
  Nucl.\ Phys.\  B {\bf 458}, 291 (1996)
  [arXiv:hep-ph/9507453].
  
  
\bibitem{Kawasaki:2007yy}
  M.~Kawasaki and K.~Nakayama,
  Phys.\ Rev.\  D {\bf 76}, 043502 (2007)
  [arXiv:0705.0079 [hep-ph]].
  
  
\bibitem{Kusenko:1997zq}
  A.~Kusenko,
  Phys.\ Lett.\  B {\bf 405}, 108 (1997)
  [arXiv:hep-ph/9704273];
  Phys.\ Lett.\  B {\bf 404}, 285 (1997)
  [arXiv:hep-th/9704073];
  S.~Kasuya and M.~Kawasaki,
  Phys.\ Rev.\  D {\bf 61}, 041301 (2000)
  [arXiv:hep-ph/9909509];
  Phys.\ Rev.\  D {\bf 62}, 023512 (2000)
  [arXiv:hep-ph/0002285].
  
  
  
 \bibitem{Kasuya:2008}
   S.~Kasuya, M.~Kawasaki and F.~Takahashi, in preparation.
 
 
\bibitem{Yamamoto:1985rd}
  K.~Yamamoto,
  Phys.\ Lett.\  B {\bf 168}, 341 (1986);
  G.~Lazarides, C.~Panagiotakopoulos and Q.~Shafi,
  Phys.\ Rev.\ Lett.\  {\bf 56}, 557 (1986).
   
  
\bibitem{Lyth:1995hj}
  D.~H.~Lyth and E.~D.~Stewart,
  Phys.\ Rev.\ Lett.\  {\bf 75}, 201 (1995)
  [arXiv:hep-ph/9502417];
  Phys.\ Rev.\  D {\bf 53}, 1784 (1996)
  [arXiv:hep-ph/9510204].
  
  
\bibitem{Choi:1996vz}
  K.~Choi, E.~J.~Chun and J.~E.~Kim,
  Phys.\ Lett.\  B {\bf 403}, 209 (1997)
  [arXiv:hep-ph/9608222];
  E.~J.~Chun, D.~Comelli and D.~H.~Lyth,
  Phys.\ Rev.\  D {\bf 62}, 095013 (2000)
  [arXiv:hep-ph/0008133];
  E.~J.~Chun, H.~B.~Kim and D.~H.~Lyth,
  Phys.\ Rev.\  D {\bf 62}, 125001 (2000)
  [arXiv:hep-ph/0008139].
  
  
   
  
\end{thebibliography}
\end{document}